\documentclass[12pt]{article}
\usepackage{amsmath}
\usepackage{amssymb}
\usepackage{epsfig}
\usepackage{euscript}
\usepackage{fancybox}
\usepackage{color}
\def\mathswitchr#1{\relax\ifmmode{\mathrm{#1}}\else$\mathrm{#1}$\fi}

\newcommand {\pslash}{\hbox{$\not\hbox{\kern-2.3pt $p$}$}}

\usepackage{cite}

\usepackage{epic}
\def\alf1{ {\alpha\over\pi} }
\begin{document}
%\input{feynman} 
%=======================================================================
\begin{titlepage}
\begin{flushright}
%{\bf MPI-PhT-2002-08}\\
{\bf BU-HEPP-09-04}\\
{\bf Jun., 2009}\\
\end{flushright}
%\vspace{0.05cm}
 
\begin{center}
{\Large On the Running of the Cosmological Constant in Quantum General Relativity}
%$^{\dagger}$}
\end{center}

\vspace{2mm}
\begin{center}
%%  {\bf   S. Jadach$^{a,b}$ and B.F.L. Ward$^{c,d}$}
{\bf   B.F.L. Ward}\\
\vspace{2mm}
%{\em $^a$CERN, Theory Division, CH-1211 Geneva 23, Switzerland,}\\
%{\em $^b$Institute of Nuclear Physics,
%        ul. Kawiory 26a, Krak\'ow, Poland,}
%{\em $^c$Werner-Heisenberg-Institut, Max-Planck-Institut fuer Physik,
%Muenchen, Germany,}\\
%{\em $^a$Werner-Heisenberg-Institut, Max-Planck-Institut fuer Physik,
%Muenchen, Germany,}\\
{\em Department of Physics,\\
 Baylor University, Waco, Texas, USA}\\
%{\em $^b$SLAC, Stanford University, Stanford, California 94309, USA,}\\
%{\em $^b$Theory Division, Saha Institute of Nuclear Physics,\\
%  Kolkata, India }\\
%{\em $^c$Department of Physics, The Citadel, Charleston, South Carolina, USA}\\
\end{center}

\vspace{5mm}
\begin{center}
{\bf   Abstract}
\end{center}
We present arguments that show what the running of the cosmological
constant means when quantum general relativity is formulated following
the prescription developed by Feynman. 
\\
\vskip 20mm
\vspace{10mm}
\renewcommand{\baselinestretch}{0.1}
\footnoterule
\noindent
%%{\footnotesize
%%\begin{itemize}
%%\item[${\dagger}$]
%%Work partly supported by US DOE grant DE-FG02-05ER41399 and 
% the Polish Government
%grants KBN 2P30225206 and 2P03B17210, the Maria Sk\l{}odowska-Curie
%Joint Fund II PAA/DOE-97-316, and
%%by NATO Grant PST.CLG.980342.
%, and by
%Polish Government grant 5P03B09320.
%%\end{itemize}
%%}
%\vspace{0.5cm}
%\begin{flushleft}
%{\bf UTHEP-00-0101}\\
%{\bf Jan, 2000}\\
%\end{flushleft}

\end{titlepage}

%=======================================================================
\def\Kmax{K_{\rm max}}\def\ieps{{i\epsilon}}\def\rQCD{{\rm QCD}}
\renewcommand{\theequation}{\arabic{equation}}
\font\fortssbx=cmssbx10 scaled \magstep2
\renewcommand\thepage{}
%\vfill\eject
\parskip.1truein\parindent=20pt\pagenumbering{arabic}\par

%\section{\bf Introduction}\label{intro}\par
Recently~\cite{attk1,sola-shp,b-s} there has been some controversy about
the meaning of a running cosmological constant in quantum field theory.
In sum in Ref.~\cite{attk1}, it is argued that the invariance of the physical vacuum energy density under renormalization group action means 
that the total response
of this quantity to a change in the renormalization scale, $\mu$, is in fact
zero, so that it does not actually run. This has been addressed in Ref.~\cite{sola-shp} by arguing that, while the total response of the vacuum energy density
to a change in such a
scale is zero, 
this still allows for that part of Einstein's theory that we ``see'' 
at low energy to contribute to the implicitly running part 
of the vacuum energy density, which is then
compensated by the dependence on the running scale 
due to both known contributions and unknown contributions from  
the possible UV completion 
of Einstein's theory.
Here, following Feynman's formulation~\cite{rpf1-2} of Einstein's theory, 
which we have recently extended to a UV finite approach~\cite{bw1} to
quantum general relativity, we will present arguments that generally agree with this latter view. Ref.~\cite{b-s} has also an equivalent view to this latter one.\par
Before proceeding any further, we need to clarify already one
important point of definition. We reference here Einstein's equation
\begin{equation}
R_{\mu\nu}-\frac{1}{2}Rg_{\mu\nu} + \Lambda g_{\mu\nu}=-8\pi G_N T_{\mu\nu}
\label{ein1}
\end{equation}
where $R_{\mu\nu}$ is the contracted Riemann tensor, $R$ is the curvature scalar, $g_{\mu\nu}$ is the metric of space-time, $G_N$ is Newton's constant,
$T_{\mu\nu}$ is the matter energy-momentum tensor and $\Lambda$ is the
cosmological constant as we will define it in our discussion.
If we take the vacuum expectation value of this equation, and follow Feynman
and expand about flat Minkowski space with the metric representation
\begin{equation}
g_{\mu\nu}=\eta_{\mu\nu}+2\kappa h_{\mu\nu}
\end{equation}
where $\kappa=\sqrt{8\pi G_N}$, $\eta_{\mu\nu}=\text{diag}(1,-1,-1,-1)$,
so that $h_{\mu\nu}$ is the quantum fluctuating field of the graviton here,
then we can move the purely gravitational contribution to the VEV,
which arises from the nonlinear part of the
``geometric'' side of Einstein's equation, to the right-hand side to get
\begin{equation}
\Lambda \eta_{\mu\nu}=-8\pi G_N <0|t_{\mu\nu}|0>
\label{cos1}
\end{equation}
where now we have defined
\begin{equation}
<0|t_{\mu\nu}|0>=<0|\left(T_{\mu\nu}+\frac{1}{\kappa^2}(R_{\mu\nu}-\frac{1}{2}Rg_{\mu\nu})|_{\text{nonlinear}}\right)|0>.
\label{cos1a}
\end{equation} 
From (\ref{cos1}), we see that, to any finite order in $\kappa$, as the 
tensor $t_{\mu\nu}$ inside the VEV operation
in (\ref{cos1a}) is conserved in the ``flat space'' sense~\cite{wein1}, 
it has zero anomalous dimension
and this proves that $\Lambda$ runs because $G_N$ runs. Indeed, that $G_N$ runs can be inferred 
immediately from the Dyson resummation
of the graviton propagator and the conservation of $t_{\mu\nu}$: in complete
analogy with QED, the ``invariant charge'' then obtains
\begin{equation}
\kappa^2(q^2)= \frac{\kappa^2}{1+\kappa^2\Pi(q^2,\mu^2,\kappa^2)}
\end{equation}
when $\kappa$ is renormalized at the point $\mu^2$ and $\Pi(q^2,\mu^2,\kappa^2)$is the respective transverse traceless renormalized proper 
graviton self-energy function.\par
The discussion in Ref.~\cite{attk1} seems to equate the discussion of the
running of the cosmological constant $\Lambda$ and the discussion
of the running of the vacuum
energy density. As defined in Einstein's equation (\ref{ein1}), the two
attendant quantities
are in fact related by a factor of $-8\pi G_N$. Since
the arguments in Ref~\cite{attk1} would 
show as well that the vacuum energy density
does not run, it would follow again that $\Lambda$ as defined here
runs with $G_N$.\par
Let us then agree to call the object analyzed in Ref.~\cite{attk1} by its
proper name the vacuum energy density $V_{\text{vac}}$. 
From (\ref{cos1}), we do
see that the relation
\begin{equation}
\Lambda =-8\pi G_N V_{\text{vac}}
\label{cos2}
\end{equation}
holds. It seems that the arguments in Ref.~\cite{sola-shp,b-s} are also
concerned with $V_{\text{vac}}$ rather than $\Lambda$
as defined in (\ref{ein1}). Accordingly, we now 
comment somewhat further with emphasis regarding the running physics
of $V_{\text{vac}}$.\par
Specifically, so far, we are working with the entire set of degrees of freedom
in Einstein's theory as formulated by Feynman, 
that is to say, we are working with
the entire set of degrees of freedom in $h_{\mu\nu}$ for example.
As Wilson has shown~\cite{wil1}, by
isolating the physics on a given scale, it is possible to 
formulate the solution of the theory in the form of scale transformations
which evolve the theory from one scale to the next, Wilsonian renormalization
group transformations. This has been pursued in Refs.~\cite{bon-reut}
in realizing Weinberg's asymptotic safety approach~\cite{wein2} 
to Einstein's theory.
If we thin the degrees of freedom a la Wilson to those relevant to a 
given scale $\mu$, as it is done in Ref.~\cite{bon-reut} for example,
then the effective action at this scale will give the equation
such as Einstein's (\ref{ein1}), with perhaps some 
higher dimensional operators added for a given level of accuracy,
but for the theory with the thinned degrees
of freedom relevant for the scale $\mu$ 
, with the attendant effective couplings at the scale $\mu$,
and this will mean that only that part
of the vacuum energy density relevant to the physics
on the scale $\mu$ will
enter into relations such as (\ref{cos1}), (\ref{cos2}). This 
means that we have
\begin{equation}
\Lambda(\mu) =-8\pi G_N(\mu) V_{\text{vac}}(\mu)
\label{cos3}
\end{equation}
following the general development of Wilson's renormalization group theory
and dropping possible irrelevant operator terms.
We conclude that both $\Lambda$ and $V_{\text{vac}}$ run when
the theory is solved a la Wilson. This is borne out by the results
in Refs.~\cite{bon-reut}. It also supports the 
arguments in Ref.~\cite{sola-shp,b-s}.\par
The basic physics underlying this running of 
both $\Lambda$ and $V_{\text{vac}}$ is as follows. If we do not thin the
degrees of freedom, we can identify a scale~\cite{sola-shp} $\mu=0$ parameter
that corresponds to the vacuum energy density of the universe
at arbitrarily long wavelengths and call that $V_{\text{vac}-\text{phys}}$
and we can then use Einstein's equation to identify $\Lambda_{\text{phys}}=-8\pi G_N(0)V_{\text{vac}-\text{phys}}$, where $G_N(0)$ is Newton's constant
at zero momentum transfer. These quantities 
$V_{\text{vac}-\text{phys}},\;\Lambda_{\text{phys}}$
would then be invariant under renormalization
and they would not run with changes in the renormalization scale $\mu$.
On the other hand, following what is done in Ref.~\cite{bon-reut} or following
Ref.~\cite{pol1} and implementing Wilsonian renormalization group theory,
we are naturally led to effective actions in which degrees of freedom have been thinned on a scale $\mu$ and the corresponding values of
$\Lambda$ and $V_{\text{vac}}$,~$\Lambda(\mu),\; V_{\text{vac}}(\mu)$, respectively, for the attendant effective action will
run with $\mu$. If a degree of freedom is integrated out of the path
integral for the theory, all of its quanta are replaced by their effects
on the remaining degrees of freedom. 
%In particular, its zero modes 
%are also integrated out. 
This means 
%that even they cannot be
%counted naively in $V_{\text{vac}}$ for the remaining degrees of freedom and
that in general $V_{\text{vac}}$ runs.
% in general\footnote{In general, on carrying out the path integral over the field components 
%which are integrated-out, one picks
%up a $\mu$-dependent constant factor that can be interpreted as 
%a part of the $V_{\text{vac}}$ for the remaining degrees of freedom, but even
%in free field theory this constant factor part is not in general equal to 
%the sum of the zero-mode contributions that the respective integrated-out
%degrees of
%freedom make to $V_{\text{vac}}$.}.
\par  
We conclude that the standard methods of the operator field do not support
the arguments in Ref.~\cite{attk1} when they are used to argue that
$\Lambda$ and $V_{\text{vac}}$ do not run. The arguments in Ref.~\cite{attk1}
regarding the renormalization invariance of $\Lambda_{\text{phys}}$ and 
$V_{\text{vac-phys}}$,
defined appropriately,
are of course correct.\par

\section*{Acknowledgments}
%The authors wish to thank JACoW for their guidance in preparing
%this template.
%
%Work supported by Department of Energy contract DE-AC02-76SF00515.
%\end{acknowledgments}
We thank Prof. J. Sola for helpful discussion. 
%for the support and kind hospitality
%of the Werner-Heisenberg Institut, MPI, Munich, while 
%a part of this work was in progress.
Work partly supported by US DOE grant DE-FG02-05ER41399 and 
% the Polish Government
%grants KBN 2P30225206 and 2P03B17210, the Maria Sk\l{}odowska-Curie
%Joint Fund II PAA/DOE-97-316, and
by NATO Grant PST.CLG.980342.\par
%\end{acknowledgments}

%%%STARTHERE
\end{document}